\documentclass[showpacs,prb,twocolumn,superscriptaddress]{revtex4}
\usepackage{epsfig}
\usepackage{bm}
\newcommand{\etal}{{\it et al.}}
\begin{document}
\title{X-ray Dichroism and the Pseudogap Phase of Cuprates}
\author{S. Di Matteo}
\affiliation{Laboratori Nazionali di Frascati INFN, via E. Fermi 40, C.P. 13,
I-00044 Frascati, Italy}
\affiliation{Equipe de physique des surfaces et interfaces, UMR-CNRS 6627 PALMS,
Universit\'e de Rennes 1, 35042 Rennes Cedex, France}
\author{M. R. Norman}
\affiliation{Materials Science Division, Argonne National Laboratory, Argonne, IL 60439, USA}
\begin{abstract}
A recent polarized x-ray absorption experiment on the high temperature cuprate
superconductor Bi$_2$Sr$_2$CaCu$_2$O$_{8+x}$ indicates the presence of broken
parity symmetry below the temperature, T*, where a pseudogap appears in
photoemission.  We critically analyze the x-ray data, and conclude that a parity-breaking
signal of the kind suggested is unlikely based on the crystal structures reported in the literature.  Possible other origins of the observed dichroism signal are discussed.  We propose x-ray
scattering experiments that can be done in order to determine whether such alternative interpretations are valid or not.
\end{abstract}

\date{\today}
\pacs{78.70.Ck, 78.70.Dm, 75.25.+z, 74.72.Hs}

\maketitle

\section{Introduction}

Twenty years since the discovery of high-temperature cuprate superconductivity, there is still no consensus on its origin.  As the field has evolved, more and more attention has been directed to the 
pseudogap region of the phase diagram in underdoped compounds and the possible relation of this phase
to the superconducting one.\cite{norman} Time-reversal breaking has been predicted to occur in this pseudogap phase due to the presence of orbital currents \cite{varma} and a subsequent experiment \cite{kaminski} using angle-dependent dichroism in photoemission has claimed to detect this.  However, this result has been challenged by others \cite{borisenko} and an independent experimental verification of this would be highly desirable. 

Recently, Kubota \etal \cite{kubota} performed Cu K edge circular and linear dichroism x-ray
experiments for underdoped Bi$_2$Sr$_2$CaCu$_2$O$_{8+x}$ (Bi2212), claiming no time-reversal breaking  of the kind predicted in Ref.~\onlinecite{varma} exists, and that, on the contrary, a parity-breaking signal (but time-reversal even) is present with the same temperature dependence as
the photoemission dichroism signal, that was interpreted as x-ray natural circular dichroism (XNCD)
as seen in other materials.\cite{alagna}

The aim of the present paper is to critically examine the conclusions of Kubota \etal \cite{kubota}  In
particular, we find that the XNCD signal for the
average \cite{average} Bi2212 crystal structure should be zero 
along all three crystallographic axes, therefore casting doubt on the original interpretation of
Ref.~\onlinecite{kubota}. To look into alternate explanations, we performed detailed numerical
simulations aimed at explaining the observed signal.
At the basis of our study is the simple observation (see, e.g., Ref.~\onlinecite{dimatoli}) that circular dichroism in absorption can be generated either by a non-magnetic effect in the electric dipole-quadrupole (E1-E2) channel (XNCD, a parity-breaking signal), or by a magnetic signal in the E1-E1 channel (parity-even).
Alternately, such a signal can be due to contamination from x-ray linear
dichroism (XNLD).
We propose x-ray experiments that could be used to investigate these matters further.

The structure of the paper is as follows.  In Sec.~II we show how symmetry constrains any possible
 XNCD signal that would be observed in Bi2212.  We also perform numerical simulations for XNCD 
assuming an alignment displaced from the c-axis, using several crystal structure refinements proposed in the literature.  We also calculate the XNLD signal and comment whether XNLD contamination could
be responsible for the observed signal.
In Sec.~III we calculate the x-ray magnetic
 circular dichroism (XMCD) signal at both the Cu K and L edges assuming magnetic order
 on either the copper or oxygen sites.  Finally, in Sec.~IV, we draw some general conclusions
 from our work. 

\section{Non-magnetic circular dichroism in Bi2212}

The structure of Bi2212 is strongly layered, with insulating BiO blocks intercalated between superconducting CuO$_2$ planes. Crystal structure refinements reveal the presence of an incommensurate modulation whose origin has been the subject of much debate.  The presence of this modulation has complicated the determination of
the average crystal structure.  Two different average structures have been proposed in the literature for Bi2212:  Bbmb \cite{bbmb,miles}  and Bb2b \cite{pet,kan,glady,glady2}, where $b$ is the modulation direction
for the superstructure. We follow the general convention in the cuprate literature and use a rotated basis compared to those in the International Tables for Crystallography \cite{inttab} (respectively, Cccm, No.~66, and Ccc2, No.~37). In this way, the $c$-crystallographic direction is orthogonal to 
the CuO$_2$ planes.

The Bbmb structure is globally centrosymmetric and, as such, does not admit a non-zero value for the parity-odd operator ${\vec L}\cdot({\hat \epsilon^*} \times {\hat \epsilon})({\vec \Omega}\cdot{\hat k})$, whose expectation value determines the XNCD signal (${\vec L}$, ${\hat k}$, ${\hat \epsilon}$ and ${\vec \Omega}$ are, respectively, the angular momentum operator, the direction and polarization of the x-ray beam, and the toroidal momentum operator, see, e.g., Refs.~\onlinecite{benoist,dimatoli}). Therefore, {\it if} the signal measured by Kubota \etal \cite{kubota} were a true XNCD signal, this would imply a lower crystal symmetry than Bbmb.  We note
that most refinements in the literature suggest Bbmb, \cite{bbmb,miles} and this crystal structure is also consistent
with recent photoemission data \cite{mans} that indicate the presence of both a glide plane and a
mirror plane based on dipole selection rules.

The other average structure that has been suggested by x-ray and neutron diffraction is Bb2b.\cite{pet,kan,glady,glady2} This space group is not centrosymmetric and, therefore, a parity breaking signal like that of XNCD is in principle allowed. However, not all wave vector directions are compatible with the presence of a XNCD signal, as demonstrated below by symmetry considerations. In the last part of Section II, we numerically calculate the XNCD for a geometrical configuration allowing a signal - like $\vec{k}\|(101)$ - by means of the multiple-scattering subroutine in the FDMNES program.\cite{yvesfdm}

In the context of this program, atomic potentials are generated using a local density approximation with a Hedin-Lundqvist form for
the exchange-correlation energy.  These potentials are then used in
a muffin tin approximation to calculate the resulting XANES signal by considering multiple scattering of the photoelectron about
the absorbing site within a one-electron approximation.\cite{natoli}  In the future, it would be desirable to repeat
these calculations by using input from self-consistent band theory, as has recently been done for the Bbmb structure
in regards to angle resolved photoemission spectra.\cite{lindroos}

In the Bb2b setting, Cu ions belong to sites of Wyckoff multiplicity $8d$. These eight equivalent copper sites can be partitioned in two groups of four sites that are related by the vector (1/2,0,1/2). Within each group the four sites are related by the symmetry operations $\{ {\hat E}, {\hat C}_{2y}, {\hat m}_{x}, {\hat m}_{z}\}$, where ${\hat E}$ is the identity, ${\hat C}_{2y}$ is a two-fold axis around the $b$ crystallographic axis, and ${\hat m}_{x(z)}$ is a mirror-symmetry plane orthogonal to the $a(c)$-axis. 
The absorption at the Cu K edge, expressed in Mbarn, can be calculated through the equations:
\begin{equation}
\sigma^{(\pm)}=\sum_{j=1}^8 \sigma_j^{(\pm)},
\label{absor1}
\end{equation}
\begin{equation}
\sigma_j^{(\pm)}=4\pi^2\alpha\hbar\omega\sum_n|\langle\Psi_n^{(j)}|{\hat O}^{(\pm)}|\Psi_0^{(j)}\rangle |^2 \delta[\hbar\omega-(E_n-E_0)]
\label{absor2}
\end{equation}

The operator ${\hat O}^{(\pm)}\equiv {\hat \epsilon}^{(\pm)}\cdot \vec{r} (1+\frac{i}{2}\vec{k}\cdot\vec{r})$ in Eq.~(\ref{absor2}) is the usual matter-radiation interaction operator expanded up to dipole (E1) and quadrupole (E2) terms, with the photon polarization ${\hat \epsilon}$
and the wave vector $\vec{k}$, where we label left- and right-handed polarization by the superscript $^{\pm}$.
$\Psi_0^{(j)}$ ($\Psi_n^{(j)}$) is the ground (excited)
state of the crystal, and $E_0$ ($E_n$) its energy.
The sum in Eq.~(\ref{absor2}) is extended over all the excited states of the system
and $\hbar\omega$ is the energy of the incoming photon, with $\alpha$ the
fine-structure constant. Finally, the index $j=1,...,8$ indicates
the lattice site of the copper photoabsorbing atom in the unit cell. Eqs.~(\ref{absor1}) and (\ref{absor2}) are the basis of the numerical calculations of the FDMNES program.\cite{yvesfdm} The eight contributions can be written as the sum of two equal parts coming from the two subsets of four ions related by the (1/2,0,1/2) translation. Within each subset, the four absorption contributions are related to one another by the symmetry operations $\sigma_2={\hat C}_{2y}\sigma_1$, $\sigma_3={\hat m}_{x}\sigma_1$, and $\sigma_4={\hat m}_{z}\sigma_1$, implying that the total absorption is:
\begin{equation}
\sigma=2 (1+{\hat m}_z)(1+{\hat m}_x) \sigma_1
\label{symmetries}
\end{equation}
The group from $\sigma_5$ to $\sigma_8$ is equivalent to the first group of four modulo a
translation (this is the reason for the factor of two in Eq.~(\ref{symmetries})). Notice that the symmetry operators in Eq.~(\ref{symmetries}) are meant to operate just on the electronic part of the operator ${\hat O}$ in Eq.~(\ref{absor2}).

\begin{figure}

\centerline{\epsfig{file=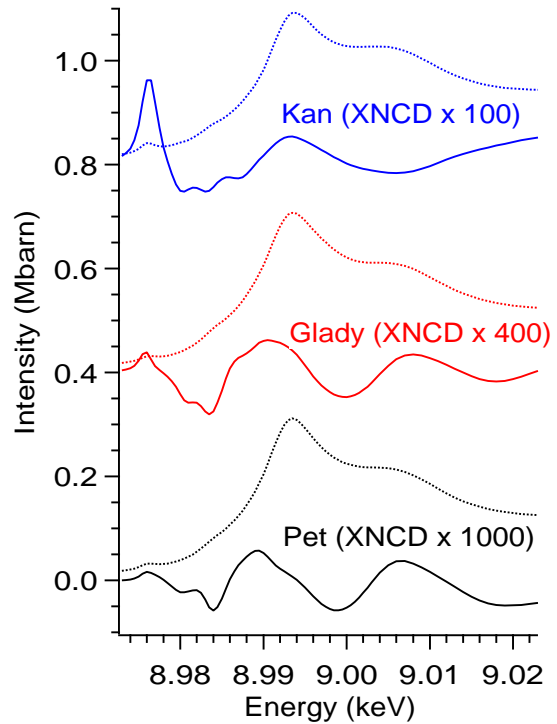,width=8.4cm}}
\caption{(Color online) XANES signal for $\vec{k} \| (001)$ and XNCD signal
for $\vec{k} \| (101)$ at the Cu K edge for three Bb2b crystal refinements, with a cluster radius of 4.9 \AA. The crystal structures are Pet for Petricek \etal \cite{pet}, Glady for Gladyshevskii and Fl\"ukiger \cite{glady}, and Kan for Kan and Moss \cite{kan}.
The XNCD signals have been multiplied by the factors indicated.
Each successive set of curves is displaced by 0.4 Mbarn.} 
\label{XNCD}

\end{figure}
 
In the case of circular dichroism, the signal is given by $\sigma=\sigma^{+}-\sigma^{-}$. If we suppose that no net magnetization is present in the material (we shall analyze the possibility of magnetism in Sec. III), then the dichroism is natural, i.e., necessarily coming from the interference E1-E2 contribution in 
Eq.~(\ref{absor2}). In this case the signal is parity-odd, which implies that ${\hat m}_{z(x)}\equiv {\hat I}{\hat C}_{2z(2x)}\rightarrow -{\hat C}_{2z(2x)}$ (${\hat I}$ is the inversion operator). Then Eq.~(\ref{symmetries}) becomes
\begin{equation}
\sigma=2 (1-{\hat C}_{2z})(1-{\hat C}_{2x}) \sigma_1
\label{symmetries2}
\end{equation}
which implies that, of the possible five second-rank tensors involved in XNCD, only the term $T^{(2)}_1-T^{(2)}_{-1}$ survives. To arrive at this result, we applied the usual operator rules on spherical tensors:\cite{varsha} ${\hat C}_{2x}T^{(2)}_m=T^{(2)}_{-m}$ and ${\hat C}_{2z}T^{(2)}_m=(-)^m T^{(2)}_{m}$. This in turn leads to a zero XNCD along the three crystallographic axes of the Bb2b crystal structure where, e.g., along the $c$-axis, the signal is proportional to $T^{(2)}_0$.

Therefore, even in the Bb2b crystal structure, the XNCD is exactly zero by symmetry when the wave vector is directed along the $c$-axis (i.e., orthogonal to the CuO$_2$ planes) as in the experiment
of Ref.~\onlinecite{kubota}. This has been further checked by numerical calculations with cluster radii up to 6.5 \AA, i.e., 93 atoms, centered on the Cu ion, based on the average crystal structures reported in
Refs.~\onlinecite{pet,kan,glady,glady2}. The only possibility to justify theoretically the experimental evidence of circular dichroism of a non-magnetic nature is either by lowering of the orthorhombic Bb2b 
symmetry, a misalignment of the ${\hat k}$-direction with respect to the $c$-axis, or contamination
from linear dichroism.  We checked all of these possibilities.

If we take into account the monoclinic supercell proposed in Ref.~\onlinecite{glady}, with space group Cc, this implies a reduction of the symmetry operations, with the loss of the two-fold screw axis. Nonetheless, the glide plane containing the normal to the CuO$_2$-planes is still present (${\hat{m}}_x$ in Eq.~\ref{symmetries}), which is responsible for the extinction rule of the quantity $\langle\Psi_n|{\vec L}\cdot({\hat \epsilon^*} \times {\hat \epsilon})({\vec \Omega}\cdot{\hat k})|\Psi_n\rangle$. Therefore the XNCD is again identically zero by symmetry, which we verified by direct numerical simulation of the supercell. Note that only a reduction to triclinic symmetry would allow for XNCD in the direction orthogonal to the CuO$_2$ planes.

\begin{figure}

\centerline{\epsfig{file=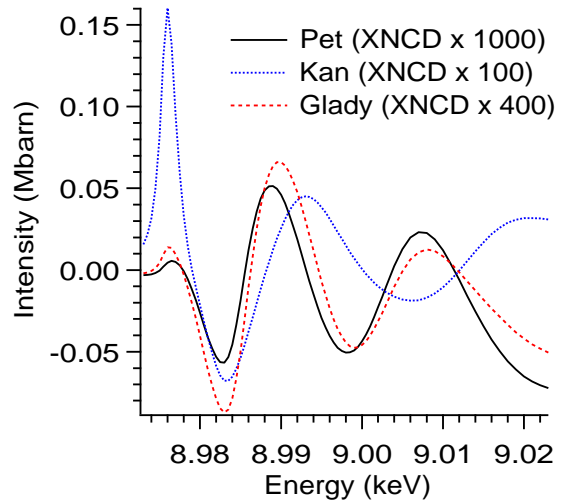,width=8.4cm}}
\caption{(Color online) XNCD signal, as in Fig.~\ref{XNCD}, but for a CuO$_5$ cluster.} 
\label{small}

\end{figure}

\begin{figure}

\centerline{\epsfig{file=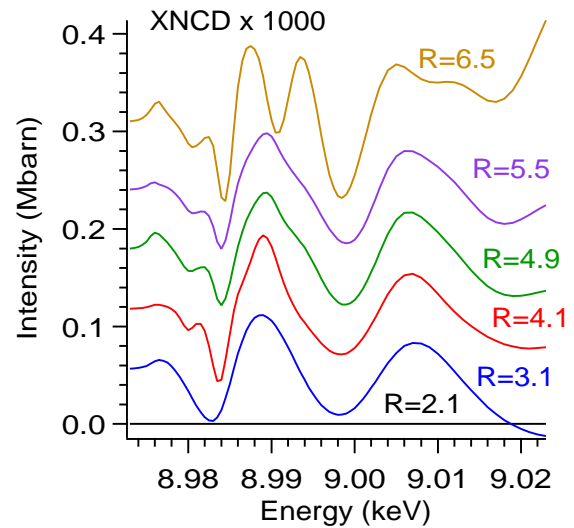,width=8.4cm}}
\caption{(Color online) XNCD signal at the Cu K edge for $\vec{k} \| (101)$ as a function of the cluster radius (in \AA) for the
crystal structure of Petricek \etal \cite{pet}.  The signal has been multiplied by 1000.
Each successive curve is displaced by 0.06 Mbarn.} 
\label{radius}

\end{figure}

We also checked for the possibility of misalignment, as shown in Fig.~\ref{XNCD}, by a direct calculation with $\vec{k} \| (101)$, corresponding to a tilting $\theta\sim 10^o$ with respect to the $c$-axis (note that the $a$ lattice constant is $\sim 5.4$ \AA, and the $c$ one $\sim 30.9$ \AA).  We first remark that the calculated XANES signal compares well to experiment (the XANES for $\vec{k} \| (001)$ and 
$\vec{k} \| (101)$ are identical on the scale of Fig.~\ref{XNCD}).  Despite this, the energy profile of the XNCD signal is different from the one reported by Kubota \etal \cite{kubota} 
The main difference is the energy extension of the calculated dichroic signal, whose oscillations persist, though with decreased intensity, more than 50 eV above the edge itself. This characteristic is present for all four Bb2b refinements we have looked at (and as well for the monoclinic supercell, which we do not show) and is at variance with the experimental results of Ref.~\onlinecite{kubota}, where the dichroic signal is confined to a small energy range around the edge.

Some comments on the calculations are in order.  Each refinement gives rise to a different XNCD signal, and their intensities are quite different as well. They are found to be strongly dependent on the magnitude of the deviation of the
atoms from their Bbmb positions, which differs significantly among the
various Bb2b refinements. 
Moreover, the structure of Kan and Moss leads to a manifestly different energy profile.
This difference is seen even for CuO$_4$ and CuO$_5$ clusters (the results for a CuO$_5$ cluster
are shown in Fig.~\ref{small})
and has to do with the large departure of this particular structure from the Bbmb one.  Although there
has been some criticism in the literature concerning this particular refinement,\cite{miles} the point
we wish to make is that each refinement has a different XNCD signal, showing how sensitive this signal
is to the actual crystal structure.   We note that Fig.~\ref{XNCD} was done for a cluster radius of 4.9 \AA, i.e., 37 atoms.  In Fig.~\ref{radius}, we show results for the refinement of Petricek \etal \cite{pet} up to a radius of 6.5 \AA{} (93 atoms),
showing the development of additional structure in the energy profile as more and more atoms are included in the
cluster.  In the real system, the effective cluster radius is limited by the photoelectron escape
depth, which is energy dependent.\cite{escape}

We also remark that the magnitude of the XNCD signal we calculated for a 10 degree 
misalignment is comparable to that measured in Ref.~\onlinecite{kubota}.
On the other hand, the signal goes as $\sin(2\theta)$, where $\theta$ measures the displacement from the $c$-axis.  We note that Kubota \etal \cite{kubota} mention that their signal was insensitive
to displacements from the $c$-axis of 5 degrees, which would argue against a misalignment given
the strong angular dependence we predict.  Moreover, we note that the size of the signal only
depends on the projection of the $k$ vector onto the $a$-$c$ plane, i.e., a signal for $\vec{k} \| (111)$
is equivalent to that for $\vec{k} \| (101)$.

The above leads us to suspect that neither misalignment, nor symmetry reduction, are the basis of the signal detected in Ref.~\onlinecite{kubota}.  We now turn to the third possibility for a non-magnetic signal, that due
to intermixing of linear dichroism.  All x-ray beams at a synchrotron have a linear polarization
component (Kubota \etal \cite{kubota} mention the possibility of up to 5\% of linear admixture).
The resulting linear dichroism, which vanishes for a uniaxial crystal,
can swamp the intrinsic XNCD signal for a 
biaxial crystal where the $a$ and $b$ directions are inequivalent (like Bi2212).  This was
shown by Goulon \etal{} for a KTiOPO$_4$ crystal with the same point group symmetry (mm2) as
Bi2212.\cite{goulonL}

\begin{figure}

\centerline{\epsfig{file=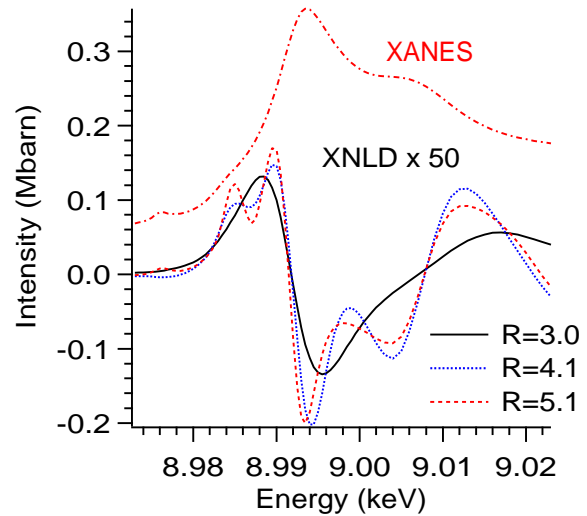,width=8.4cm}}
\caption{(Color online) XNLD signal at the Cu K edge for $\vec{k} \| (001)$ as a function of the cluster radius (in \AA) for the
crystal structure of Gladyshevskii \etal \cite{glady2}.  The XNLD signal has been multiplied by 50.
The XANES curve has been displaced by 0.05 Mbarn.} 
\label{XNLD}

\end{figure}

To analyze this further, we show the linear dichroism (the XANES signal for $\vec{E} \| (010)$ minus
the one for $\vec{E} \| (100)$), calculated for the Bb2b refinement of Gladyshevskii \etal, \cite{glady2} for various
cluster radii, in Fig.~\ref{XNLD}.  Similar results have been obtained for the other crystal structures,
including the Bbmb refinement of Miles \etal \cite{miles}

The energy profile, with a positive peak followed by a negative peak, and its
location at the absorption edge, is very reminiscent of the data.  Moreover, the size of the XNLD signal
is large, meaning that only a few percent admixture is necessary to explain the size of the signal
seen in Ref.~\onlinecite{kubota}.  One issue is that Kubota \etal \cite{kubota} did report the existence of an XNLD signal,
but also claim that it is temperature independent.  This is somewhat puzzling, as there are significant
changes of the lattice constants with temperature.\cite{milesT}  One obvious question would be why
such an XNLD contamination would only appear below T*, though it should be remarked that there are
anomalies in the superstructure periodicity near T*.\cite{milesT}  A definitive test would be
to rotate the sample under the beam, as any XNLD signal would vary as $\cos(2\phi)$ where
$\phi$ is the in-plane angle relative to the $b$-axis.
Any circular dichroism (XNCD or XMCD) is instead $\phi$-independent.

A final possibility would be a small energy shift between the left and right circularly polarized beams.  Differentiating the
absorption edge in Fig.~\ref{XNCD} would indeed lead to a signal similar to that seen in
Ref.~\onlinecite{kubota} (but with an enhanced positive peak relative to the negative peak).
But such an energy shift is difficult to imagine with the particular experimental set up
used.

The observed dichroism signal as a function of energy is also reminiscent of that typically seen for {\it magnetic} circular dichroism: in this case the signal would be of E1-E1 origin and its main features are indeed expected to be at the edge itself.  In addition, the nature of the observed signal (a single sharp positive peak followed by a single sharp negative peak) is also much like a magnetic signal where the main features are expected to be more localized in energy starting from the rising edge of the
absorption.
Whether this possibility is realistic or not can only be determined by a quantitative calculation, which
we offer in the next section.

\section{Magnetic dichroism in Bi2212}

Over the years,
there have been several claims of possible magnetic order in the pseudogap phase of cuprate superconductors.
Recently, a magnetic signal at a (101) Bragg vector has been observed below T* for several underdoped YBa$_2$Cu$_3$O$_{6+x}$ (YBCO) samples by polarized neutron diffraction.\cite{bourges}  The signal, corresponding to a moment
of order 0.05-0.1 $\mu_B$, is not
simple ferromagnetism as it was not observed at the (002) Bragg vector.  Even more recently, a Kerr rotation below T* has been detected in underdoped YBCO corresponding
to a net ferromagnetic moment of 10$^{-4}$ $\mu_B$.\cite{kapitulnik}

This motivates us to consider the possibility of a magnetic origin for the x-ray circular dichroism detected in Ref.~\onlinecite{kubota}. We restate that in absorption (see, e.g., Ref.~\onlinecite{dimatoli}), circular dichroism can be generated either by a non-magnetic effect in the E1-E2 channel (XNCD, a parity-breaking signal), or by a magnetic signal in the E1-E1 channel (XMCD, parity-even). The first possibility, analyzed in the previous section, does not seem to be compatible with the experimental results. In order to analyze the second possibility, we need to provide the lattice with a magnetic structure that has a net magnetization (otherwise, the XMCD is zero). In what follows, we shall suppose two magnetic distributions, the first with the magnetic moments on the Cu sites, the second on the planar O sites.
The numerical calculations are performed with the relativistic extension of the multiple-scattering program in the FDMNES code,\cite{yvesfdm} and provide results that are an extension of those of the previous section.

\begin{figure}

\centerline{\epsfig{file=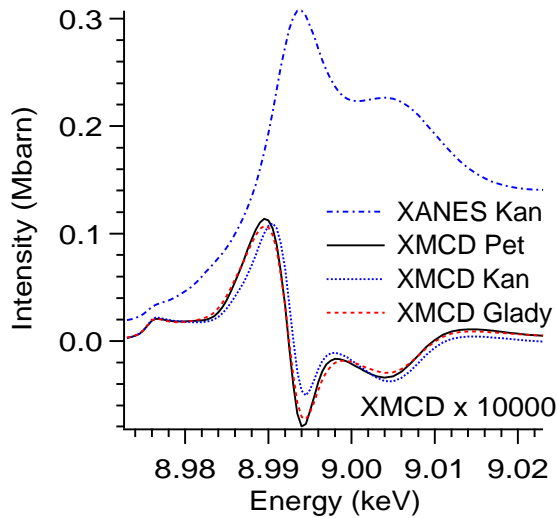,width=8.4cm}}
\caption{(Color online) XMCD for $\vec{k} \| (001)$ at the Cu K edge with a ferromagnetic moment of 0.1 $\mu_B$ along the c-axis at each copper site. The cluster radius is 4.1 \AA. The XMCD signal has been multiplied by 10000.  The crystal structures are those of Fig.~\ref{XNCD}.} 
\label{XMCD}

\end{figure}

\begin{figure}

\centerline{\epsfig{file=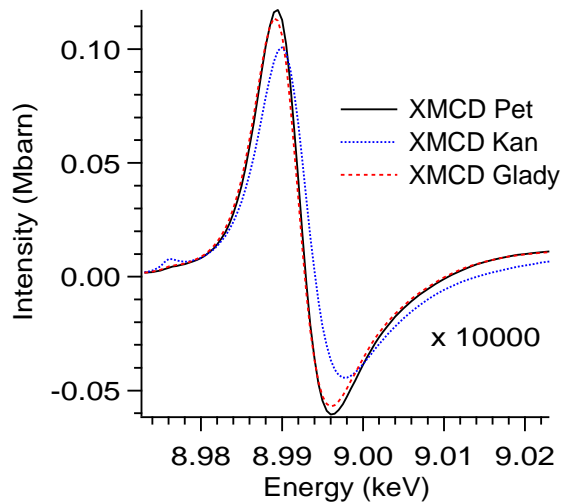,width=8.4cm}}
\caption{(Color online) XMCD as in Fig.~\ref{XMCD}, but with a cluster radius of 3.1 \AA, i.e., a CuO$_5$ cluster.} 
\label{XMsmall}

\end{figure}

\begin{figure}

\centerline{\epsfig{file=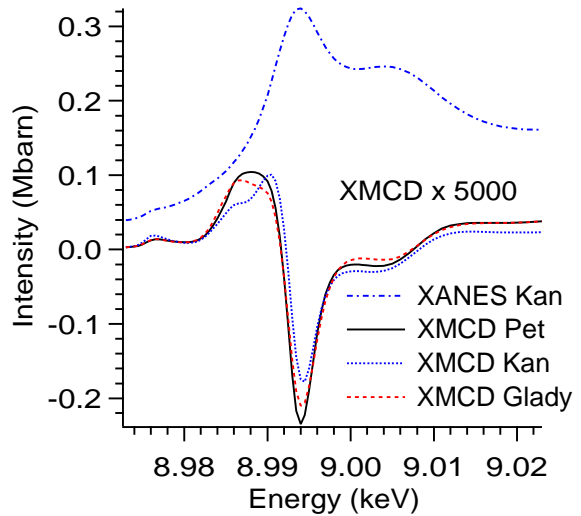,width=8.4cm}}
\caption{(Color online) XMCD for $\vec{k}\|(001)$ at the Cu K edge with a ferromagnetic moment of 0.1 $\mu_B$ at each planar oxygen site. The XMCD signal has been multiplied by 5000.  The XANES signal has been displaced by 0.02 Mbarn.} 
\label{oxygen}

\end{figure}

The details of the calculations are as follows: we used again the average crystal structures
discussed in Section II. For each magnetic configuration we employed clusters with radii ranging
from 3.1 \AA{} (a CuO$_5$-cluster), to 4.9 \AA{} (37 atoms) around the Cu photoabsorbing ion. In the first set of calculations, shown in Figs.~\ref{XMCD} and \ref{XMsmall}, we built the input potential from a magnetic configuration of 4.55 $3d_{\uparrow}$ electrons and 4.45 $3d_{\downarrow}$ electrons (i.e., a moment of 0.1 $\mu_B$ per copper site).  In the second set of calculations, shown in Fig.~\ref{oxygen}, we built the input potential from a magnetic configuration of 2.05 $2p_{\uparrow}$ electrons and 1.95 $2p_{\downarrow}$ electrons (i.e., a moment of 0.1 $\mu_B$ per planar oxygen site).
The following results are noteworthy: 

$a)$ Differently from the XNCD calculations shown in Fig.~\ref{XNCD}, all crystal structures give basically the same XMCD spectra. The reason for this behavior may be related to the fact that XMCD, when x-rays are orthogonal to the CuO$_2$ planes, mainly depends on the in-plane magnetisation density and on the in-plane crystal structure, which is quite similar for the various refinements. 

$b)$ There is a more marked dependence on the cluster radius compared to the XNCD, as shown by the comparison of Fig.~\ref{XMCD} and Fig.~\ref{XMsmall}. The calculations with a radius bigger than 4.1 \AA { }(6 Cu, 9 O, 4 Sr, and 4 Ca), above the pre-edge energy, are basically equivalent to those shown in Fig.~\ref{XMCD}, with a positive peak at the edge energy, followed by a double negative peak, the
latter at variance with the experimental results. On the contrary, the energy shape obtained for a radius of
3.1 \AA {} is very close to the experimental one, with a single negative peak after the sharp positive one, with relatively good agreement in the energy position and width. We could be tempted to suppose, therefore, that the virtual photoelectron has a very small mean free path before decaying and it is sensitive just to the nearest neighbor oxygens. Indeed we checked that an identical XMCD profile is obtained with just the in-plane CuO$_4$-cluster.  On the other hand, the size of the signal we calculate
is about an order of magnitude smaller than that seen in Ref.~\onlinecite{kubota}.  Since the XMCD signal is proportional to
the moment, then we would need a moment of $\sim$1 Bohr magneton per copper to
have a comparable signal.  Such a huge moment would have been observed previously by neutron scattering if it existed. Of course we cannot exclude that spurious effects, such as strain fields, could have influenced the measurement.

$c)$ The energy profile in the case of magnetization at the oxygen sites is not much different from the copper case, except for the deeper negative peak around E $\sim 8.994$ keV, as shown in Fig.~\ref{oxygen}. Also in this case the different crystal structure refinements give basically equivalent results, as again the CuO$_2$ planes are practically equivalent in the various cases. Note that the relative intensity is equivalent to the copper case, as 0.1 Bohr magnetons per planar oxygen corresponds to 0.2 Bohr magnetons per CuO$_2$ cell (note we multiply by 5000 in Fig.~\ref{oxygen} as compared to 10000
in Fig.~\ref{XMCD}).

\begin{figure}

\centerline{\epsfig{file=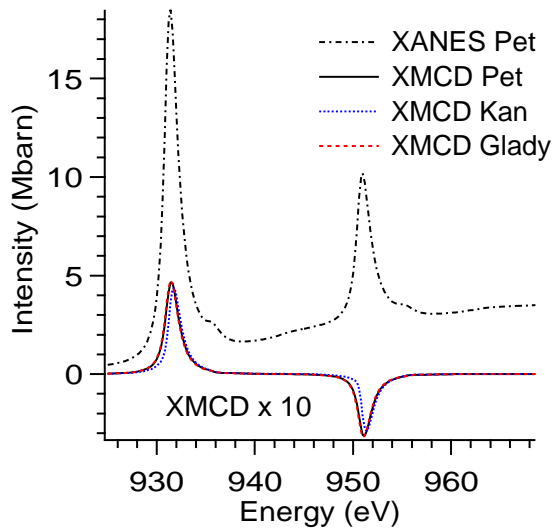,width=8.4cm}}
\caption{(Color online) XMCD for $\vec{k} \| (001)$ at the Cu L$_{2,3}$ edges with a ferromagnetic moment of 0.1 $\mu_B$ along the c-axis at each copper site. The cluster radius is  4.1 \AA. The XMCD signal has been multiplied by 10.} 
\label{Ledge}

\end{figure}

We also performed simulations for the Cu L$_{2,3}$ edges \cite{foot1} for the magnetic configuration corresponding to Fig.~\ref{XMCD}, as shown in Fig.~\ref{Ledge}, which can be compared with future experimental investigations in order to confirm whether or not a net magnetization exists in this compound. 

Finally, we remark that the dependence of the XMCD signal on the tilting angle $\theta$ (i.e., the displacement of the photon wave vector from the $c$-axis) goes like $\cos(\theta)$ and therefore the signal is not very sensitive to small displacements of 5 degrees, as noted by Kubota \etal \cite{kubota} This different angular dependence from the XNCD signal suggests a relatively easy way to unravel the question experimentally: it is sufficient to measure the $\theta$ (azimuthal) dependence of the signal, noting that any XNLD contamination would be tested by the $\phi$ (polar) dependence of the signal.

\section{Conclusions}

In our opinion, the experiments of Kubota \etal \cite{kubota} have raised more questions than they have answered. Although not treated in our paper, we believe that their results from the measurement of non-reciprocal linear dichroism are at this stage not conclusive, as only one direction for the toroidal moment has been investigated, of the two possible suggested by the orbital current pattern of Varma.\cite{varma} The analysis performed in Section II showed, moreover, that the claimed XNCD signal is probably unjustified. In fact, even though the space group Bb2b is non-centrosymmetric, XNCD is absent by symmetry when the x-ray wave vector is chosen orthogonal to the CuO$_2$ planes, as in the
experimental measurement geometry of Ref.~\onlinecite{kubota}. The same extinction rule survives for the monoclinic supercell structure refined in Ref.~\onlinecite{glady}. Moreover, in both cases, it seems hard to mantain the hypothesis of misalignment, due to the experimental localization of the energy profile around the main absorption edge, which is absent in the calculations.
We also note the difference of XNCD from the photoemission dichroism results of Kaminski \etal \cite{kaminski} A direct comparison is however not immediate, as the former represents a  $\hat{q}$-integrated version of the latter (here $\hat{q}$ is the solid angle in the space of the photoelectron wave-vector, see, e.g., Ref. \onlinecite{dimatoli}).  A more
likely explanation is an XNLD contamination (Fig.~\ref{XNLD}), but then the challenge is to
understand why such an effect would only exist below T*.
We note that an optics experiment for an optimal doped Bi2212 sample has seen a change in 
linear birefrigence below T$_c$, which was accompanied by a non-zero circular
birefringence.\cite{koby}  In addition, both Bi2223 \cite{flukiger} and the Fe analogue of 
Bi2212 \cite{lepage} exhibit supercells with 222 space groups which would allow for  dichroism.
So, it is conceivable that there is a subtle structural transition associated with T* which we suggest
could be looked for by diffraction experiments.

A final comment about the physical quantities detected by x-ray circular dichroism, either natural or magnetic, is in order.  
At the K edge of transition metal oxides, XMCD in the E1-E1 channel, at the excitation energy ${\cal E}=\hbar\omega-E_{\rm edge}$, gives information on the expectation value of ${\hat L}\cdot ({\hat \epsilon^*} \times {\hat \epsilon})$ for the excited states at the energy ${\cal E}$. The orbital angular momentum is either induced from a spin moment via spin-orbit coupling (as calculated here) or directly by an orbital
current (as in the scenario advocated in Ref.~\onlinecite{varma} \cite{foot3}). No direct spin information is available at the K edge and therefore an XMCD measurement is not directly related to the ground state magnetic moment as for the L edge. Moreover, the observed states are those with $p$-like angular momentum projection on the photoabsorbing Cu ion, which are extended and therefore mainly sensitive to the influence of the oxygen atoms surrounding the Cu site. In this case, the main contribution to the XMCD energy profile is  expected in an energy range of 10-20 eV from the main edge to the first shoulder in the XANES spectrum, as found in Ref.~\onlinecite{kubota} and in our own XMCD calculations.

Although the results of Section III are in principle consistent with Ref.~\onlinecite{kubota}, the size
of the ferromagnetic moment necessary to get a signal of the magnitude seen by experiment, $\sim$1 Bohr magneton, is excessive.  If such a large ferromagnetic moment existed, it would have surely been seen by neutron scattering.  From this point of view, experiments performed at the Cu L edge and O K edge would be desirable as
they are more sensitive to the presence of a magnetic moment.

Finally, we would like to remark that XNCD along the c-axis is insensitive to orbital currents. These
latter, confined to the CuO$_2$ planes, develop a parity breaking characterized
by a toroidal moment ($\vec{\Omega}$) within the CuO$_2$ planes. The XNCD experiment
of Ref.~\onlinecite{kubota} would only be sensitive to the projection of the toroidal
moment out of this plane (i.e., along the direction of the x-ray wavevector).
Therefore, if performed as stated, it cannot tell us about any possible orbital current order.

To conclude, we believe that the various interpretations, XNCD, XNLD, or XMCD, have their drawbacks, and therefore the origin of the experimental signal of Ref.~\onlinecite{kubota} is still open. In this sense, further experimental checks of the energy extension of the dichroic signal would be highly desirable.
Based on our results, the most stringent experimental test on the physical origin of the signal would
come from the measurement of the dependence on the tilting (azimuthal) angle $\theta$, due to the different dependences of XNCD and XMCD, as well as the dependence on the in-plane (polar) 
angle $\phi$, which would test for any possible XNLD contamination.

\section{Acknowledgments}

The authors thank John Freeland, Zahir Islam, Stephan Rosenkranz, Daniel Haskel
and Matti Lindroos for various discussions.
This work is supported by the U.S. DOE, Office of Science, under Contract No.~DE-AC02-06CH11357.
SDM would like to thank the kind hospitality of the ID20 beamline staff at the ESRF.


\begin{thebibliography}{99}
\bibitem{average}
The average crystal structure is the base orthorhombic unit cell before the incommensurate 
superstructure is taken into account.  As the latter
involves a translation operator, it should not affect the symmetry arguments in this paper.
\bibitem{escape}
The results presented involve a convolution of the calculated spectrum with 
both a core hole (1.9 eV for the Cu K edge) and a photoelectron
lifetime, with the latter having a strong energy dependence (ranging up to 15 eV with a
midpoint value at 30 eV above the Fermi energy).  But the calculation has a fixed cluster radius.
\bibitem{foot1}
It should be remarked, though, that single-particle based approaches can miss some features of the spectral weight transfer between the $L_2$ and $L_3$ edges.
\end{thebibliography}
\end{document}